\documentclass{acm_proc_article-sp}
\pdfoutput=1
\makeatletter
\let\@copyrightspace\relax
\makeatother

\usepackage{amsmath}
\usepackage{amssymb}
\usepackage{graphicx}
\usepackage{subfig}
\usepackage{keyval}
\usepackage{everysel}
\usepackage{ragged2e}
\usepackage{verbatim}
\usepackage{array}
\usepackage{afterpage}
\usepackage{url}
\usepackage{flushend}
\usepackage{multirow}
\usepackage{enumitem}

\def\sharedaffiliation{
\end{tabular}
\begin{tabular}{c}}

\begin{document}

\title{Investigating Keyphrase Indexing with Text Denoising}

\numberofauthors{2}
\author{
\alignauthor
Rushdi Shams\\
       \email{rshams@csd.uwo.ca}
\alignauthor
Robert E. Mercer\\
       \email{mercer@csd.uwo.ca}
\sharedaffiliation
				\affaddr{Department of Computer Science}\\
        \affaddr{University of Western Ontario}\\
        \affaddr{London, ON N6A 5B7, Canada}
}

\maketitle
%%%%%%%%%%%%%%%%%%%%%%%%%%%%%%%%%%%%%%%%%%%%%%%%%%%%%%%%%%%%%%%%%%%%%%%%%%%%%%%%%%%%%%%%%%%%%%%%%%%%%%%%%%%%%%%%%%%%%%%%%%%%%%%%%%%%%%%%%%%%%%%%%%%%%%%%%%%%%%%%
\begin{abstract}
In this paper, we report on indexing performance by a state-of-the-art keyphrase indexer, Maui, when paired with a text extraction procedure called \textit{text denoising}. Text denoising is a method that extracts the denoised text, comprising the content-rich sentences, from full texts. The performance of the keyphrase indexer is demonstrated on three standard corpora collected from three domains, namely food and agriculture, high energy physics, and biomedical science. Maui is trained using the full texts and denoised texts.  The indexer, using its trained models, then extracts keyphrases from test sets comprising full texts, and their denoised and noise parts (i.e., the part of texts that remains after denoising). Experimental findings show that against a gold standard, the denoised-text-trained indexer indexing full texts, performs either better than or as good as its benchmark performance produced by a full-text-trained indexer indexing full texts.
\end{abstract}
%%%%%%%%%%%%%%%%%%%%%%%%%%%%%%%%%%%%%%%%%%%%%%%%%%%%%%%%%%%%%%%%%%%%%%%%%%%%%%%%%%%%%%%%%%%%%%%%%%%%%%%%%%%%%%%%%%%%%%%%%%%%%%%%%%%%%%%%%%%%%%%%%%%%%%%%%%%%%%%%
\category{H.3.1}{Information Storage and Retrieval}{Content Analysis and Indexing}[Indexing method]
\category{H.3.3}{Information Storage and Retrieval}{Information Search and Retrieval}
\category{H.3.4}{Information Storage and Retrieval}{Systems and Software}[Performance evaluation (efficiency and effectiveness)]
%%%%%%%%%%%%%%%%%%%%%%%%%%%%%%%%%%%%%%%%%%%%%%%%%%%%%%%%%%%%%%%%%%%%%%%%%%%%%%%%%%%%%%%%%%%%%%%%%%%%%%%%%%%%%%%%%%%%%%%%%%%%%%%%%%%%%%%%%%%%%%%%%%%%%%%%%%%%%%%%
\terms{Experimentation, Performance.}
%%%%%%%%%%%%%%%%%%%%%%%%%%%%%%%%%%%%%%%%%%%%%%%%%%%%%%%%%%%%%%%%%%%%%%%%%%%%%%%%%%%%%%%%%%%%%%%%%%%%%%%%%%%%%%%%%%%%%%%%%%%%%%%%%%%%%%%%%%%%%%%%%%%%%%%%%%%%%%%%
\keywords{Keyphrase extraction, topic extraction, indexing, text denoising, keyphrase indexer, machine learning model, fog index} % NOT required for Proceedings
%%%%%%%%%%%%%%%%%%%%%%%%%%%%%%%%%%%%%%%%%%%%%%%%%%%%%%%%%%%%%%%%%%%%%%%%%%%%%%%%%%%%%%%%%%%%%%%%%%%%%%%%%%%%%%%%%%%%%%%%%%%%%%%%%%%%%%%%%%%%%%%%%%%%%%%%%%%%%%%%

%%%%%%%%%%%%%%%%%%%%%%%%%%%%%%%%%%%%%%%%%%%%%%%%%%%%%%%%%%%%%%%%%%%%%%%%%%%%%%%%%%%%%%%%%%%%%%%%%%%%%%%%%%%%%%%%%%%%%%%%%%%%%%%%%%%%%%%%%%%%%%%%%%%%%%%%%%%%%%%%
\section{Introduction}
\label{introduction}
%%%%%%%%%%%%%%%%%%%%%%%%%%%%%%%%%%%%%%%%%%%%%%%%%%%%%%%%%%%%%%%%%%%%%%%%%%%%%%%%%%%%%%%%%%%%%%%%%%%%%%%%%%%%%%%%%%%%%%%%%%%%%%%%%%%%%%%%%%%%%%%%%%%%%%%%%%%%%%%%
Since they provide high-level descriptions of document contents, keyphrases serve as the meta-descriptions as well as a means to effective document retrieval from digital libraries. Other reasons to use keyphrases include but are not limited to document similarity measure, classification and clustering, topic search, web tag clouds and document summarization~\cite{Witten-et-al:1999}. 

Today, automatic keyphrase indexing is a popular notion which eliminates several drawbacks of manual indexing such as conflicting time and effort, and poor choice of keyphrases. Among the automatic keyphrase indexers, several are tested across domains~\cite{Witten-et-al:1999}\cite{Frantzi-et-al:1998}\cite{Olena:2009}\cite{Medelyan:2008}\cite{Turney:2000}\cite{Witten-et-al:2004} while many are domain-specific~\cite{Aronson-et-al:2004}. Most of these indexers are trained with full documents using algorithms like Na\"{i}ve Bayes and Bagging to extract keyphrases from full-text test documents. A revealing experiment by Witten \textit{et al.}~\shortcite{Witten-et-al:1999} demonstrates that the performance of the indexers depends not only on these features but also on document size. As they apply their full-text trained Keyphrase Extraction Algorithm (hereinafter, KEA) on paper abstracts and compare against a gold standard, they find its performance on these reduced texts somewhat inferior and not competitive to that on full texts. The authors concluded that this anomaly was unequivocal as fewer author-assigned keyphrases appear in the chosen reduced texts than in the entire document.

\textit{Text Denoising} is a method proposed by Shams and Mercer~\shortcite{Shams:2011} which reduces the amount of text in biomedical papers to 30\% of the original. This 30\% of the text is selected based on the Fog Index readability score \cite{Fog:1969} and is called \textit{denoised text}; the remaining text is called the \textit{noise text}. In this introductory work denoised text is shown to be the more content-rich portion of the full text as it contains most of the biomedical concepts that are explicitly or implicitly connected with biomedical relations. Although tests have been carried out only with biomedical research articles, the authors conclude that Fog Index can be a useful indicator of content richness for other genres and different purposes although the threshold of 30\% might need to be reconsidered. %Getting to know that and the effect of document size on indexing performances, it is interesting to see if text denoising can be an effective means for improved indexing even with a reduced set of texts.
%%%%THIS SENTENCE DIDN'T MAKE SENSE TO ME. YOU SEEM TO WANT TO USE IT AS A MOTIVATION FOR THE PAPER, BUT IT NEEDS REWORDING IF YOU WANT TO KEEP IT.

In this paper, we report on the performance of a state-of-the-art keyphrase indexer named Maui \cite{Olena:2009} when paired with text denoising. We use three standard full text corpora from the food and agriculture, high energy physics, and biomedical science domains. From each corpus, we develop training sets comprising full texts and their denoised parts. The test sets are composed of full texts, and their denoised and noise parts. For training and testing each dataset, we use a standard 10-fold cross validation. We show experimentally that although a threshold of 30\% performs well for biomedical relation extraction, it is 70\% for keyphrase indexing. To evaluate Maui, we use quantitative measures like precision, recall and F-score as well as qualitative measures like inter-indexer agreements. Experimental results show that Maui, with denoised texts, performs either better or comparably to its benchmark performance---those with full-text trained models to extract keyphrases from full-text test sets. 

The remainder of this paper provides background on text denoising and the keyphrase indexer, Maui (Section \ref{background}), and discusses the methods for training and testing the indexer (Section \ref{methods}), an analysis of the results (Section \ref{results}), and ends with some concluding remarks (Section \ref{conclusions}).

%%%%%%%%%%%%%%%%%%%%%%%%%%%%%%%%%%%%%%%%%%%%%%%%%%%%%%%%%%%%%%%%%%%%%%%%%%%%%%%%%%%%%%%%%%%%%%%%%%%%%%%%%%%%%%%%%%%%%%%%%%%%%%%%%%%%%%%%%%%%%%%%%%%%%%%%%%%%%%%%
\section{Background}
\label{background}
%%%%%%%%%%%%%%%%%%%%%%%%%%%%%%%%%%%%%%%%%%%%%%%%%%%%%%%%%%%%%%%%%%%%%%%%%%%%%%%%%%%%%%%%%%%%%%%%%%%%%%%%%%%%%%%%%%%%%%%%%%%%%%%%%%%%%%%%%%%%%%%%%%%%%%%%%%%%%%%%
%Section preamble
In this section, we briefly discuss the text denoising method as well as Maui, the keyphrase indexer. 

%%%%%%%%%%%%%%%%%%%%%%%%%%%%%%%%%%%%%%%%%%%%%%%%%%%%%%%%%%%%%%%%%%%%%%%%%%%%%%%%%%%%%%%%%%%%%%%%%%%%%%%%%%%%%%%%%%%%%%%%%%%%%%%%%%%%%%%%%%%%%%%%%%%%%%%%%%%%%%%%
\subsection{Text Denoising}
%%%%%%%%%%%%%%%%%%%%%%%%%%%%%%%%%%%%%%%%%%%%%%%%%%%%%%%%%%%%%%%%%%%%%%%%%%%%%%%%%%%%%%%%%%%%%%%%%%%%%%%%%%%%%%%%%%%%%%%%%%%%%%%%%%%%%%%%%%%%%%%%%%%%%%%%%%%%%%%%
%How text denoising works
In their paper, Witten \textit{et al.}~\cite{Witten-et-al:1999} have demonstrated that the performance of the indexer KEA has been reduced when extracting keyphrases from paper abstracts. Similarly, the performance of biomedical relation miners that attempt to extract relations among drugs, chemicals, diseases, genes and proteins from paper abstracts is such that a number of biomedical ontologies like OMIM (Online Mendelian Inheritance in Man) and GO (Gene Ontology) use human annotators to extract relations from full texts. This procedure is time-consuming as well as error-prone. To overcome these shortcomings, Shams and Mercer~\shortcite{Shams:2011} proposed a method that identifies those areas within a text, called denoised text, where content information, such as biomedical relations, is more likely to occur. The authors suggested that the describing of biomedical relations lengthens sentences and increases the use of polysyllabic words. Some readability indexes, the Fog Index \cite{Fog:1969} in particular, are based on these two factors. They proceeded to use Fog Index to measure sentence readability and showed experimentally that the 30\% of the sentences which had the lowest-readability, the denoised part of a text, contained the relations of interest.

\begin{figure}[hb]
\begin{center}
\includegraphics[scale=0.45]{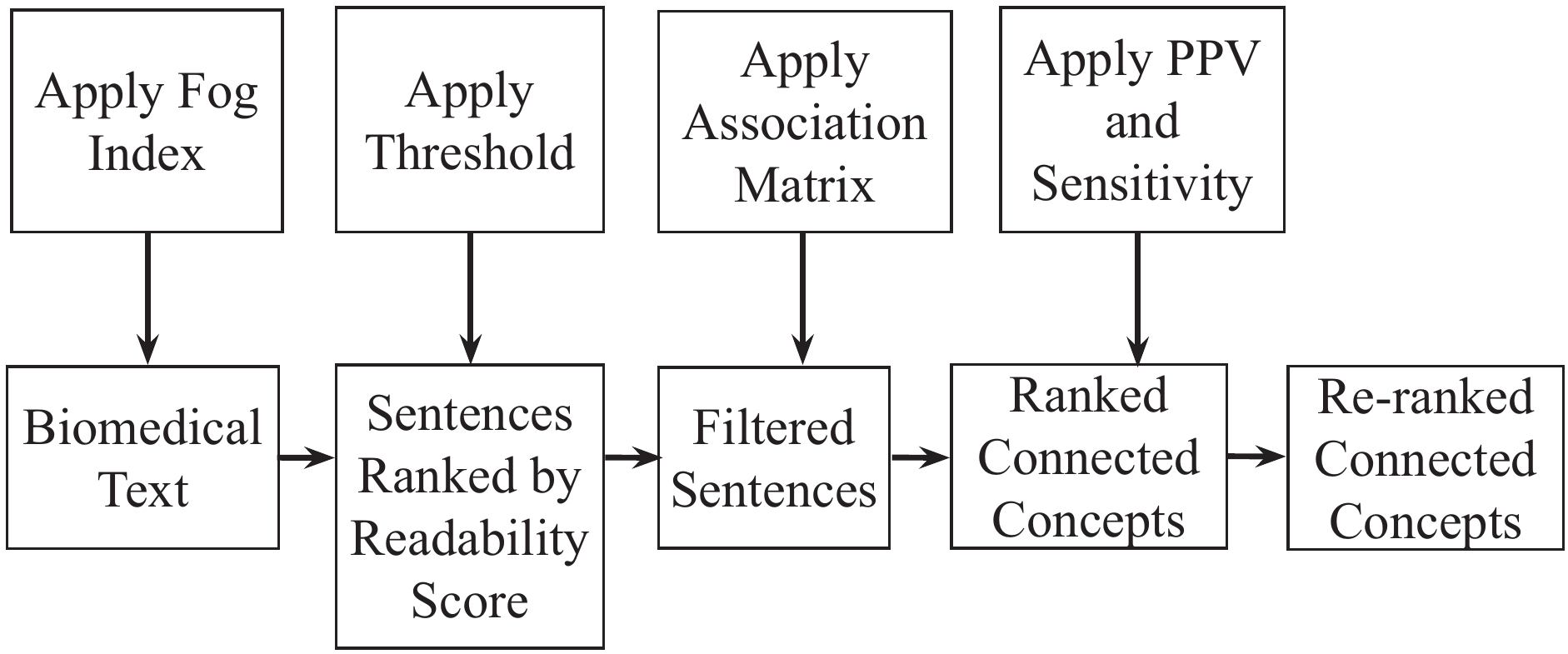}
\end{center}
\caption{Text denoising and connected concept extraction method described by Shams and Mercer \cite{Shams:2011}}
\label{textdenoising}
\end{figure}

Figure \ref{textdenoising} illustrates the text denoising method applied to biomedical texts.
%How text denoising performs
Text Denoising has been evaluated with a corpus comprising 24 full texts that describe four related pairs of disease and chemical components. This method extracted pairs of biomedical concepts from the denoised part of the texts of which about 75 percent are reported as related according to the Unified Medical Language System's (UMLS) semantic relations network. It is noteworthy that the rest of the text, called noise text, did not contain any related biomedical concepts of interest.

%End of Text Denoising

%%%%%%%%%%%%%%%%%%%%%%%%%%%%%%%%%%%%%%%%%%%%%%%%%%%%%%%%%%%%%%%%%%%%%%%%%%%%%%%%%%%%%%%%%%%%%%%%%%%%%%%%%%%%%%%%%%%%%%%%%%%%%%%%%%%%%%%%%%%%%%%%%%%%%%%%%%%%%%%%
\subsection{Maui}
%%%%%%%%%%%%%%%%%%%%%%%%%%%%%%%%%%%%%%%%%%%%%%%%%%%%%%%%%%%%%%%%%%%%%%%%%%%%%%%%%%%%%%%%%%%%%%%%%%%%%%%%%%%%%%%%%%%%%%%%%%%%%%%%%%%%%%%%%%%%%%%%%%%%%%%%%%%%%%%%
Automatic keyphrase indexing has been in practice for half a century~\cite{Luhn:1959}, but until recently the performance was not notable.
Maui\footnote{http://code.google.com/p/maui-indexer/} \cite{Olena:2009} is the final successor of a legacy of keyphrase indexers and inherits from and builds upon both of its predecessors KEA \cite{Witten-et-al:1999}\cite{Witten-et-al:2004} and KEA++\footnote{http://www.nzdl.org/Kea/}~\cite{Medelyan:2008}. Maui uses $13$ features (among them are \textit{tf}$\times$\textit{idf} and first occurrence (i.e., the number of words preceding a keyphrase normalized by the total number of words in a document) inherited from KEA; node degree (i.e., the number of connections between a candidate phrase and the other candidates in the SKOS hierarchy) and keyphrase length inherited from KEA++) to develop machine learning models and extract keyphrases. Maui's own features include \textit{tf}, \textit{idf}, last occurrence, spreads (i.e., number of words between first and last occurrence of a phrase), semantic relatedness and generality (i.e., position in the vocabulary hierarchy). 

Furthermore, Maui uses a feature named keyphraseness (i.e., the probability that a candidate phrase is present in the training documents) for domain-specific keyphrase extraction. Frank \textit{et al.}~\cite{Witten-et-al:1999} reported that the use of this feature increases indexers' performances if their training and test documents are retrieved from the same domain. Maui, moreover, can be incorporated with any domain-specific controlled vocabulary written in SKOS\footnote{http://www.w3.org/2001/sw/wiki/SKOS/Datasets} (Simple Knowledge Organization System) hierarchical format. Besides domain-specific keyphrase indexing, Maui has the capability for both free-text indexing and indexing for any domain that lacks a controlled vocabulary. For the latter case, Maui uses Wikipedia as a domain-independent controlled vocabulary. 

Maui has been evaluated with texts from three different domains: food and agriculture, nuclear physics, and biomedical science. Although the performance is not comparable to that of English, it has also been tested with texts written in French and Spanish. Experimental outcomes show that Maui outperforms its predecessors in both cases \cite{Olena:2009}. Moreover, it performs significantly better than Medical Text Indexer (MTI\footnote{http://skr.nlm.nih.gov/SKR\_API/index.shtml})~\cite{Aronson-et-al:2004} and BibClassify\footnote{http://invenio-demo.cern.ch/help/hacking/bibclassify-admin-guide}~\cite{Pepe:2008} for indexing biomedical and physics texts, respectively.

\begin{figure}[hbt]
\noindent 
\begin{centering}
\includegraphics[scale=0.45]{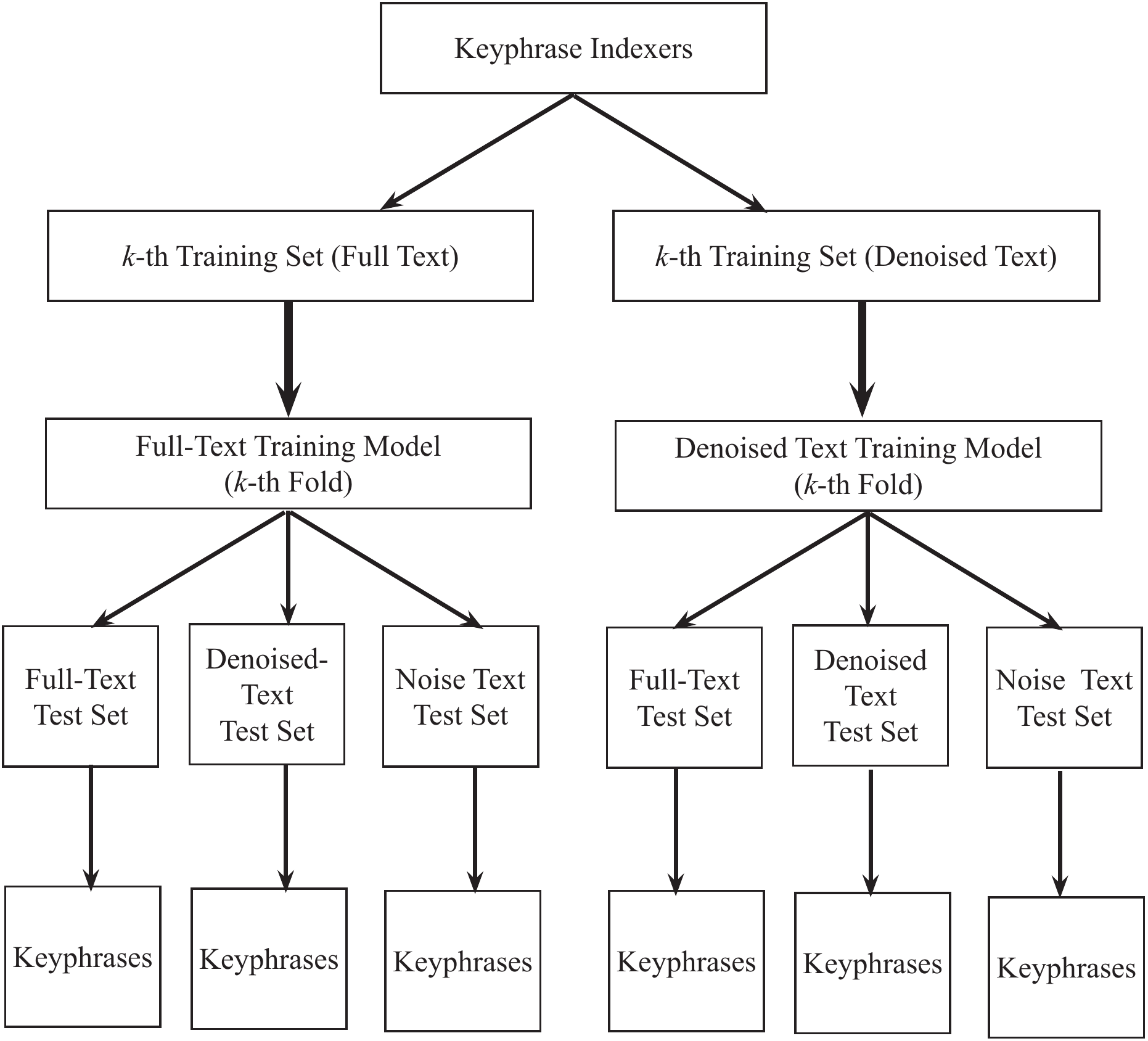}
\par
\end{centering}
\caption{Keyphrase extraction from \textit{k}-th fold}
\label{fig:methods}
\end{figure}

%Describes the methodology of the research
%%%%%%%%%%%%%%%%%%%%%%%%%%%%%%%%%%%%%%%%%%%%%%%%%%%%%%%%%%%%%%%%%%%%%%%%%%%%%%%%%%%%%%%%%%%%%%%%%%%%%%%%%%%%%%%%%%%%%%%%%%%%%%%%%%%%%%%%%%%%%%%%%%%%%%%%%%%%%%%%
\section{Methodology}
\label{methods}
%%%%%%%%%%%%%%%%%%%%%%%%%%%%%%%%%%%%%%%%%%%%%%%%%%%%%%%%%%%%%%%%%%%%%%%%%%%%%%%%%%%%%%%%%%%%%%%%%%%%%%%%%%%%%%%%%%%%%%%%%%%%%%%%%%%%%%%%%%%%%%%%%%%%%%%%%%%%%%%%
%Section preamble
In this section, we describe the datasets, training and testing procedure, performance measures, and the means to find the appropriate text denoising threshold for keyphrase indexing. 

We use the datasets and follow the experimental protocols set by Medelyan~\cite{Olena:2009} except that we train Maui not only on full texts but also on their denoised parts and test it on full texts as well as their denoised and noise parts. 

%Describes the dataset
%%%%%%%%%%%%%%%%%%%%%%%%%%%%%%%%%%%%%%%%%%%%%%%%%%%%%%%%%%%%%%%%%%%%%%%%%%%%%%%%%%%%%%%%%%%%%%%%%%%%%%%%%%%%%%%%%%%%%%%%%%%%%%%%%%%%%%%%%%%%%%%%%%%%%%%%%%%%%%%%
\subsection{Datasets}
%%%%%%%%%%%%%%%%%%%%%%%%%%%%%%%%%%%%%%%%%%%%%%%%%%%%%%%%%%%%%%%%%%%%%%%%%%%%%%%%%%%%%%%%%%%%%%%%%%%%%%%%%%%%%%%%%%%%%%%%%%%%%%%%%%%%%%%%%%%%%%%%%%%%%%%%%%%%%%%%

In this experiment, to train and test Maui, we use three standard corpora of full texts and keyphrases associated with them from three different domains. These corpora were collected by Medelyan~\cite{Olena:2009} during her doctoral research. 

The first dataset, which is referred to as FAO-780, contains $780$ full-text documents and their keyphrases. The documents have been selected randomly from the Food and Agriculture Organization (FAO) data repository. The dataset contains about $24$ million words ($30,800$ words on average per document) and $6,225$ keyphrases ($8$ keyphrases on average per document, ranging from $2$ to $23$). An in-depth analysis of the documents reveals that the set is composed of both research articles and experimental reports. 

Our second dataset comprises $290$ full-text documents and their keyphrases on high energy physics randomly collected from the European Organization for Nuclear Research (abbreviated as CERN) document server and thus named CERN-290. Each document contained therein has an average of $6,300$ words and $7$ keyphrases. CERN-290 is the smallest dataset that we use in this experiment and contains mainly experimental reports. 

We also used the NLM-500 corpus, collected by the NLM Indexing Initiative \cite{aronson:2000} during the development of MTI, which consists of $500$ biomedical research articles. This corpus has documents with average length of $4,500$ words and an average number of assigned keyphrases of $15$ ranging from $2$ to $30$. The contents of the dataset are mainly scholarly research articles collected from the National Library of Medicine repository.

%%%%%%%%%%%%%%%%%%%%%%%%%%%%%%%%%%%%%%%%%%%%%%%%%%%%%%%%%%%%%%%%%%%%%%%%%%%%%%%%%%%%%%%%%%%%%%%%%%%%%%%%%%%%%%%%%%%%%%%%%%%%%%%%%%%%%%%%%%%%%%%%%%%%%%%%%%%%%%%%
\subsection{Training and Testing}
\label {train-test}
%%%%%%%%%%%%%%%%%%%%%%%%%%%%%%%%%%%%%%%%%%%%%%%%%%%%%%%%%%%%%%%%%%%%%%%%%%%%%%%%%%%%%%%%%%%%%%%%%%%%%%%%%%%%%%%%%%%%%%%%%%%%%%%%%%%%%%%%%%%%%%%%%%%%%%%%%%%%%%%%
In our first attempts at pairing Maui with reduced texts, we noted that Witten \textit{et al.}~\cite{Witten-et-al:2004}, using the Computer Science Technical Reports (CSTR) corpus, showed that any training set containing more than $25$ documents has very little effect on the indexer's performance. In our initial experiments with Maui we followed this protocol by randomly choosing $25$ training and $100$ test documents from the NLM-500 corpus. Against a gold standard---author-assigned keyphrases for the $100$ test documents---we measured Maui's precision, recall and F-score. From these experiments, we have seen that

\begin{itemize}[leftmargin=*]
\item although the performance of Maui with the denoised text trained model is better than that with full-text trained model, the improvement is not statistically significant and the improvement does not reflect on the entire population, and
\item Maui's performance improves if we increase the text denoising threshold from 30\% to 40\%.
\end{itemize}

The first observation indicates that the methods followed by Witten \textit{et al.}~\cite{Witten-et-al:2004} can be effective for certain domains but are not an effective means for many others while the latter indicates that for keyphrase indexing, the text denoising threshold is not 30\%.

Therefore, we decided to use a more conventional \textit{k}-fold experimental approach. We followed the experimental procedure illustrated in Figure \ref{fig:methods}. We consider the full texts from each dataset and divide them randomly into 10 equal-sized folds where the documents in one fold do not overlap with the others. In addition, we keep the denoised and noise parts of each fold. Then, we apply a standard 10-fold cross validation to train and test Maui. To generate each pair, we keep one of the $10$ folds out as our testing set and combine the rest of the $9$ folds as our training set. Doing this $10$ times, each time leaving out a different one from the $10$ folds as a testing set, we get $10$ pairs. We train Maui on the training sets comprising full texts and denoised texts from each fold. In this way, we develop $20$ trained models for the entire $10$ folds. The models the indexers develop from full texts are called \textit{full-text trained models} and those that are developed from denoised texts are called \textit{denoised text trained models}.

As the trained models are created, the indexers then apply them, \textit{k}-th full-text trained model and \textit{k}-th denoised text trained model, to extract keyphrases from the \textit{k}-th test set composed of full texts, and their denoised and noise parts. According to the average number of keyphrases in every document, we had the indexers extract $8$ keyphrases, $7$ keyphrases, and $15$ keyphrases for each document in the FAO-780, CERN-290, and NLM-500 test sets, respectively. The extracted keyphrases are then compared against a gold standard which are the author assigned keyphrases associated with the test documents. The testing has been carried out for the rest of the folds and the performance measures described in section \ref{performance} are then averaged.

It is noteworthy that during training and testing, we used controlled vocabularies for the respective domains, and during testing we set the minimum and maximum length of the keyphrases to be extracted per document to $1$ and $5$, respectively, as this is the default setting of Maui.

\begin{figure*}[bth]
  \centering
  \subfloat[Error rates for different denoising thresholds with denoised texts]{\label{fig:faothreshold1}\includegraphics[width=0.45\textwidth]{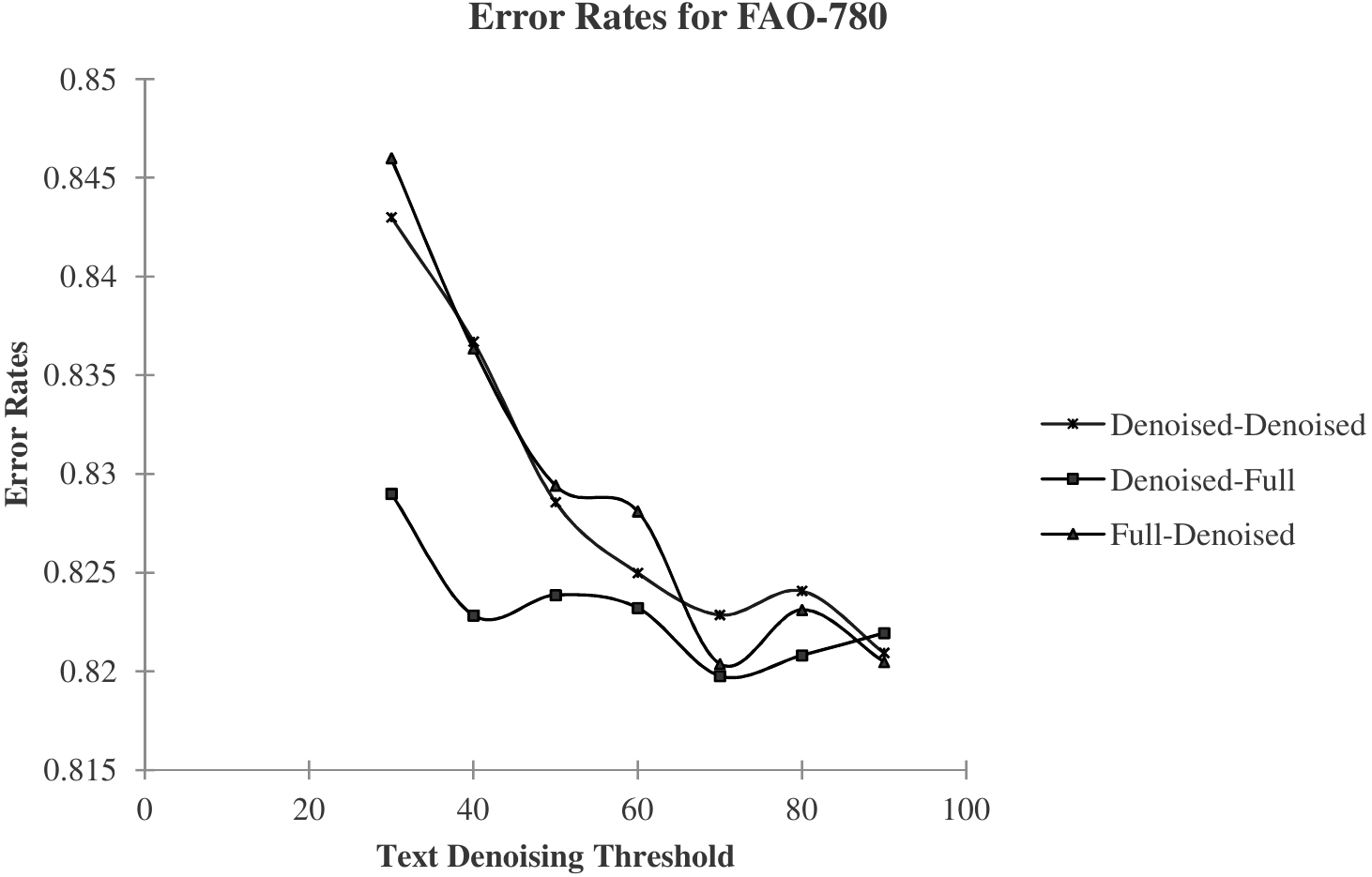}}      \hspace{1cm}          
  \subfloat[Error rates for different denoising thresholds with noise texts]{\label{fig:faothreshold2}\includegraphics[width=0.45\textwidth]{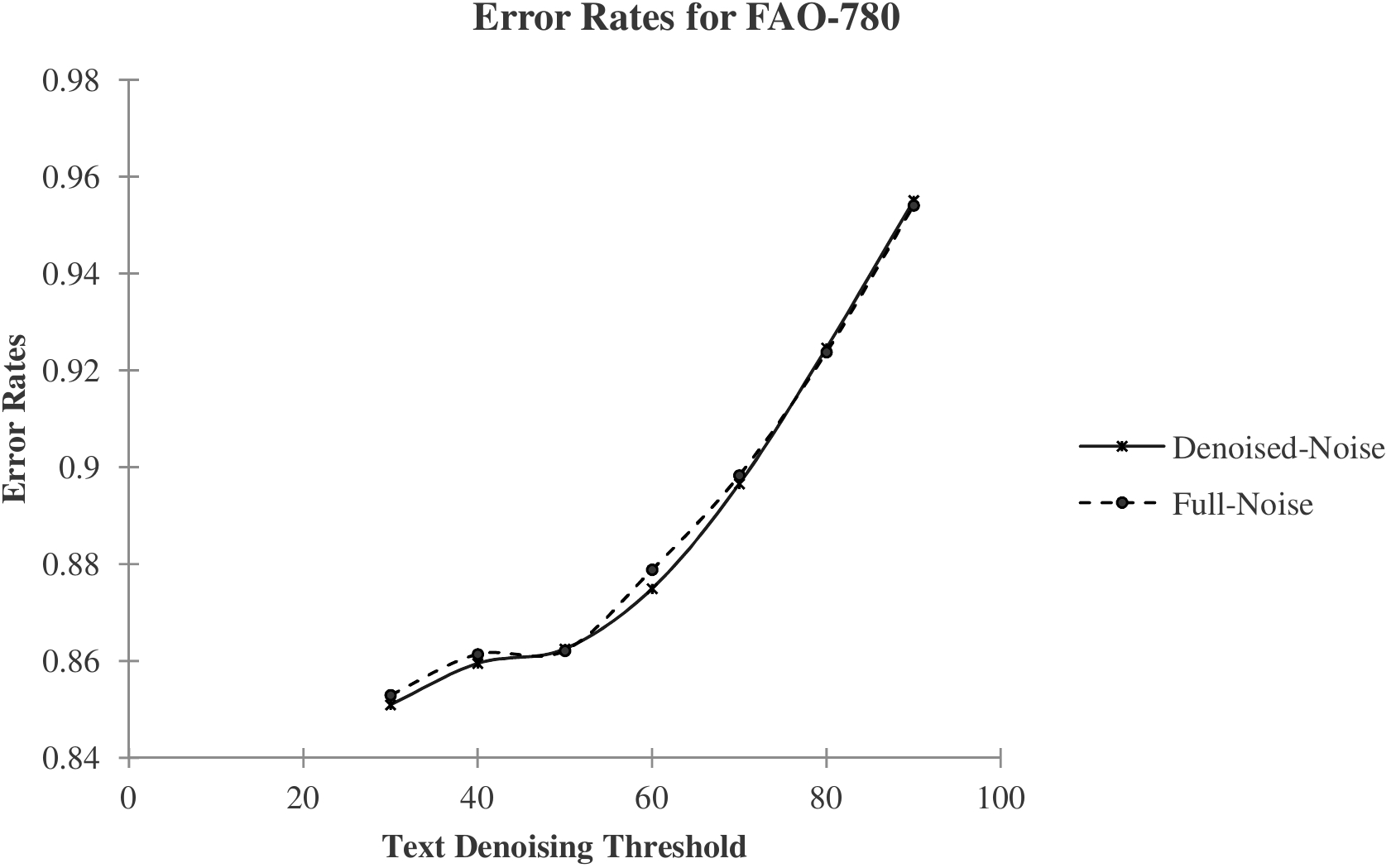}}
  \caption{Text denoising threshold for FAO-780 dataset}
  \label{fig:faothreshold}
\end{figure*}

%%%%%%%%%%%%%%%%%%%%%%%%%%%%%%%%%%%%%%%%%%%%%%%%%%%%%%%%%%%%%%%%%%%%%%%%%%%%%%%%%%%%%%%%%%%%%%%%%%%%%%%%%%%%%%%%%%%%%%%%%%%%%%%%%%%%%%%%%%%%%%%%%%%%%%%%%%%%%%%%
\subsection{Performance Measures}
\label{performance}
%%%%%%%%%%%%%%%%%%%%%%%%%%%%%%%%%%%%%%%%%%%%%%%%%%%%%%%%%%%%%%%%%%%%%%%%%%%%%%%%%%%%%%%%%%%%%%%%%%%%%%%%%%%%%%%%%%%%%%%%%%%%%%%%%%%%%%%%%%%%%%%%%%%%%%%%%%%%%%%%
In this experiment, we use the conventional quantitative measures for performance evaluation---precision, recall and F-score. In addition, we use three inter-indexing agreement measures popularly used for qualitative indexing assessment~\cite{Medelyan:2006}. The measures are called Hooper's (\textit{H}), Roll-ing's (\textit{R}) and Cosine (\textit{C}) inter-indexing agreements. The common property of these agreement measures is that they provide the number of correct keyphrases in relation to the size of the two sets of keyphrases being compared. We briefly summarize these agreement measures for the reader's convenience. If \textit{M} and \textit{N} are the number of idiosyncratic keyphrases assigned by two indexers and \textit{O} is the number of phrases two indexers have in common, then Hooper's measure~\cite{Hooper:1965} is
\begin{displaymath}H(\textit{indexer}_{1},\textit{indexer}_{2})=\frac{O}{M+N-O}.\end{displaymath}
Similarly, Rolling's measure~\cite{Rolling:1981} is defined as
\begin{displaymath}R(\textit{indexer}_{1},\textit{indexer}_{2})=\frac{2{\cdot}O}{M+N}.\end{displaymath}
Cosine measure uses the geometric mean instead of Rolling's arithmetic mean. Thus, Cosine measure can be written as
\begin{displaymath}C (\textit{indexer}_{1},\textit{indexer}_{2})=\frac{O}{\sqrt{M{\cdot}N}}. \end{displaymath}
The last two measures are almost identical unless the sets radically vary. It can be noted that Hooper's and Rolling's measures are identical to Jaccard's coefficient and the Dice coefficient, respectively, which are used to measure statistical similarity between two sets. The closer the agreement measures are to 1, the more the indexers agree on extracted keyphrases.

We also calculate the error rates for every cross validation. The error rate is defined as
\begin{displaymath}E=\frac{\textit{FP}+\textit{FN}}{N},\end{displaymath}
where \textit{FP} and \textit{FN} are the number of false positives and false negatives, respectively and \textit{N} is the total number of instances in the test sets. The reasons for using the error rates are twofold, first, to find a text denoising threshold described in Section \ref{threshold} and second, to measure a \textit{10-fold cross validated paired t-test} \cite{parsons} to report significant improvement of Maui when paired with denoised texts. For any two given sets of results, we consider their error rates to calculate a paired \textit{t}-value. If this calculated \textit{t}-value lies outside $\pm 2.26$ with a degree of freedom $9$, then the difference between the set whose results have the lower error rate and the other set is said to be statistically significant at significance level $\alpha$ = $0.05$.

\begin{figure*}[tbh]
  \centering
  \subfloat[Error rates for different denoising thresholds with denoised texts]{\label{fig:cernthreshold1}\includegraphics[width=0.45\textwidth]{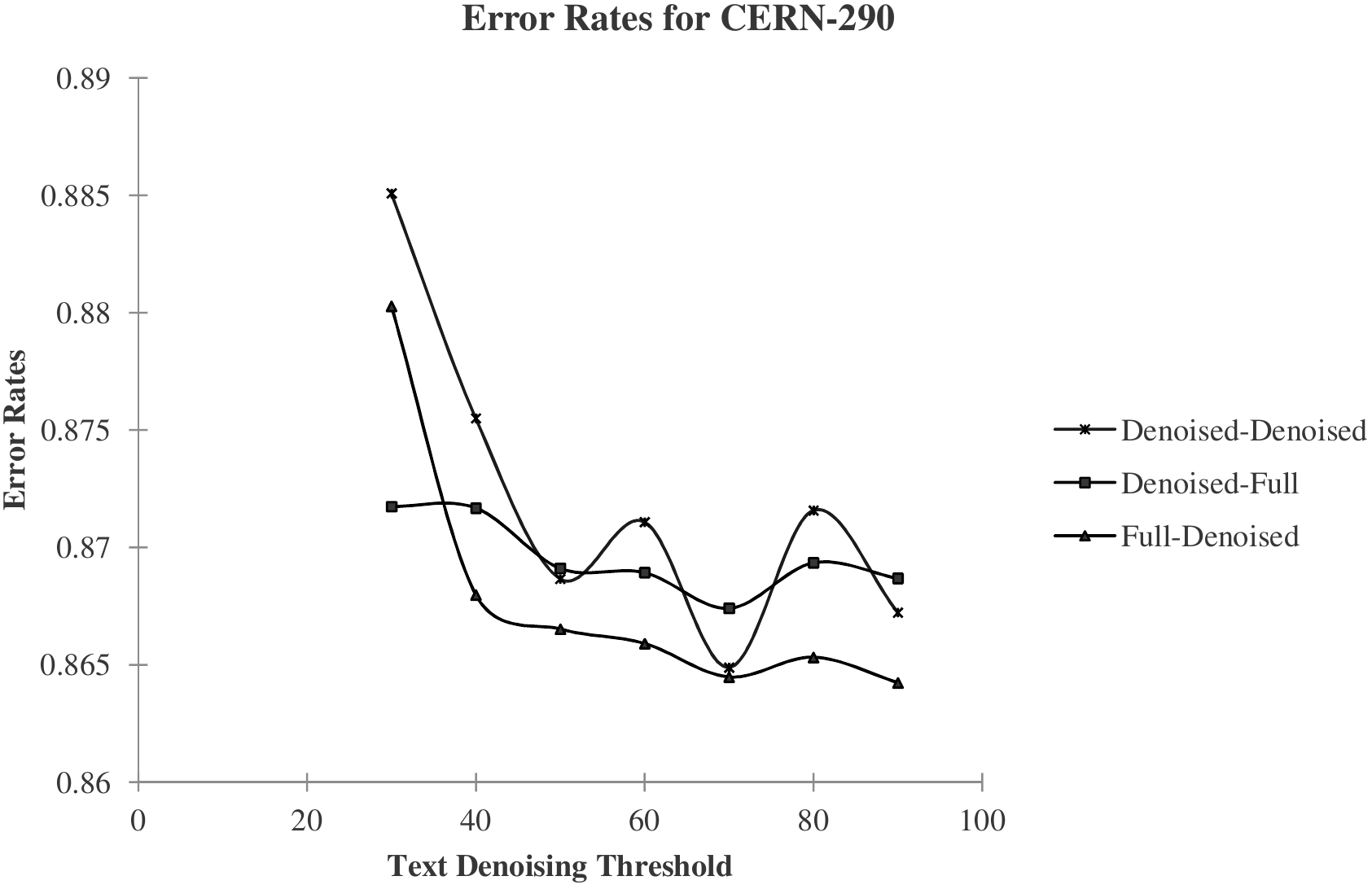}}      \hspace{1cm}          
  \subfloat[Error rates for different denoising thresholds with noise texts]{\label{fig:cernthreshold2}\includegraphics[width=0.45\textwidth]{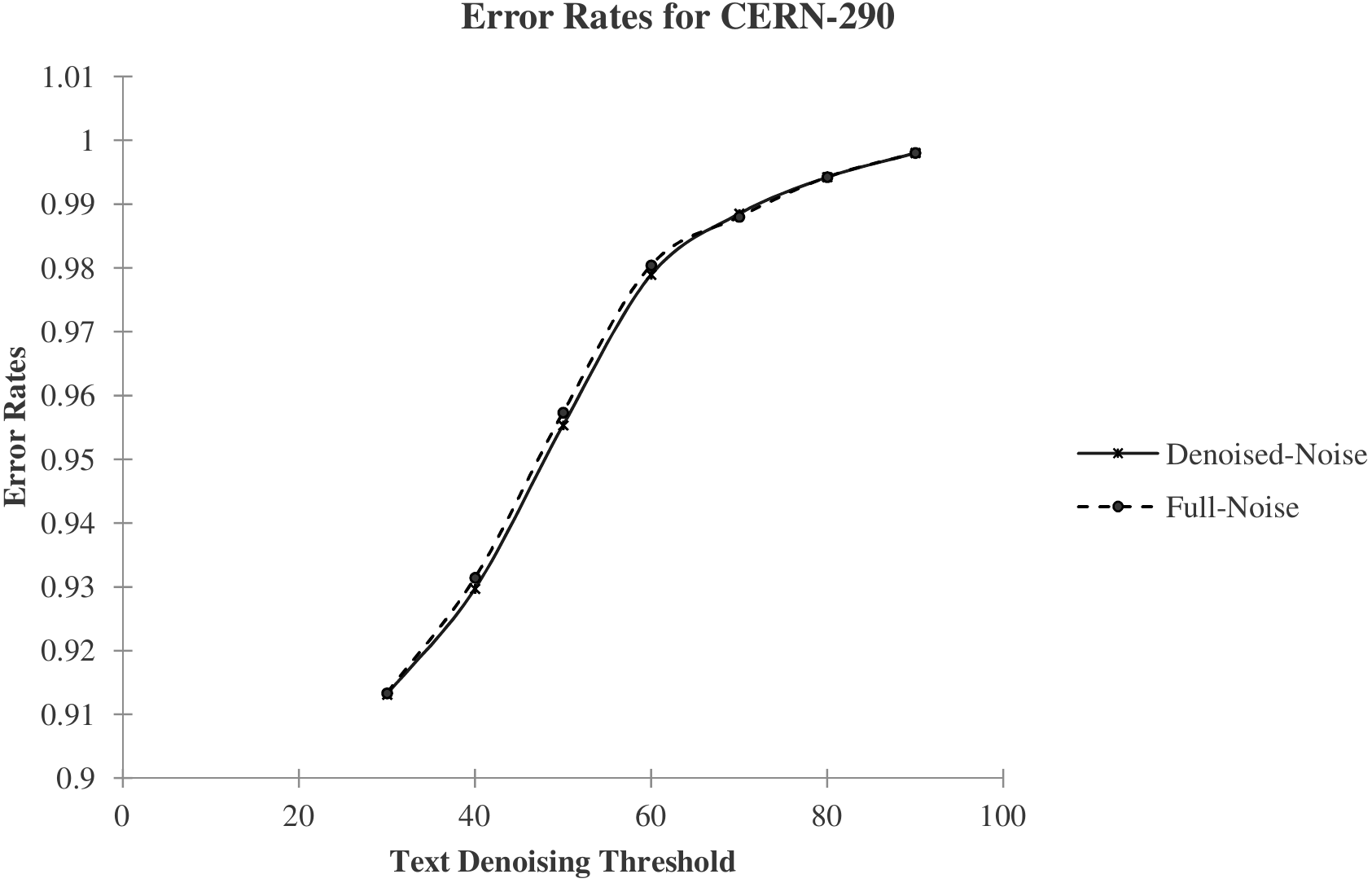}}
  \caption{Text denoising threshold for CERN-290 dataset}
  \label{fig:cernthreshold}
\end{figure*}

\begin{figure*}[tbh]
  \centering
  \subfloat[Error rates for different denoising thresholds with denoised texts]{\label{fig:nlmthreshold1}\includegraphics[width=0.45\textwidth]{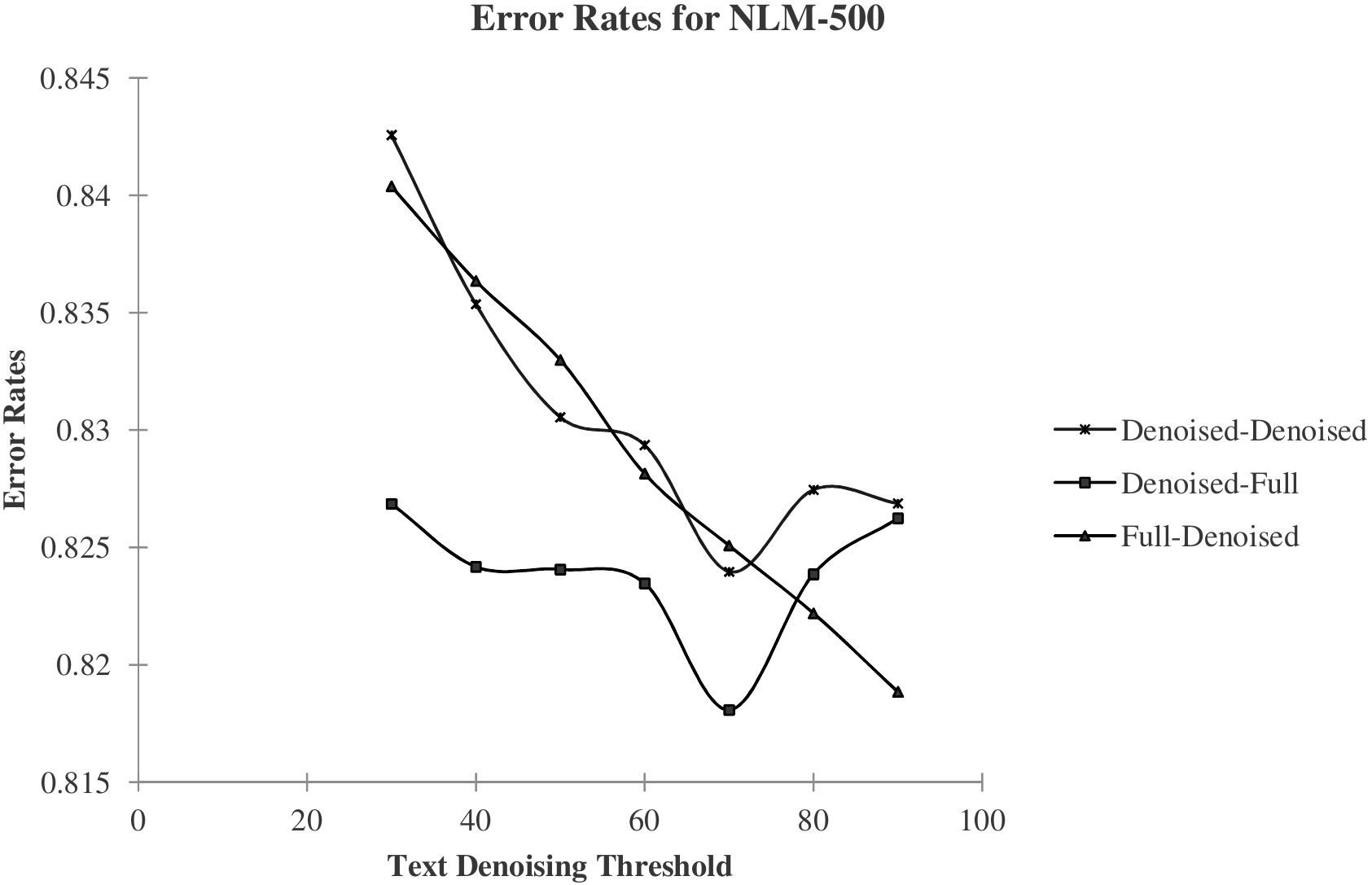}}      \hspace{1cm}          
  \subfloat[Error rates for different denoising thresholds with noise texts]{\label{fig:nlmthreshold2}\includegraphics[width=0.45\textwidth]{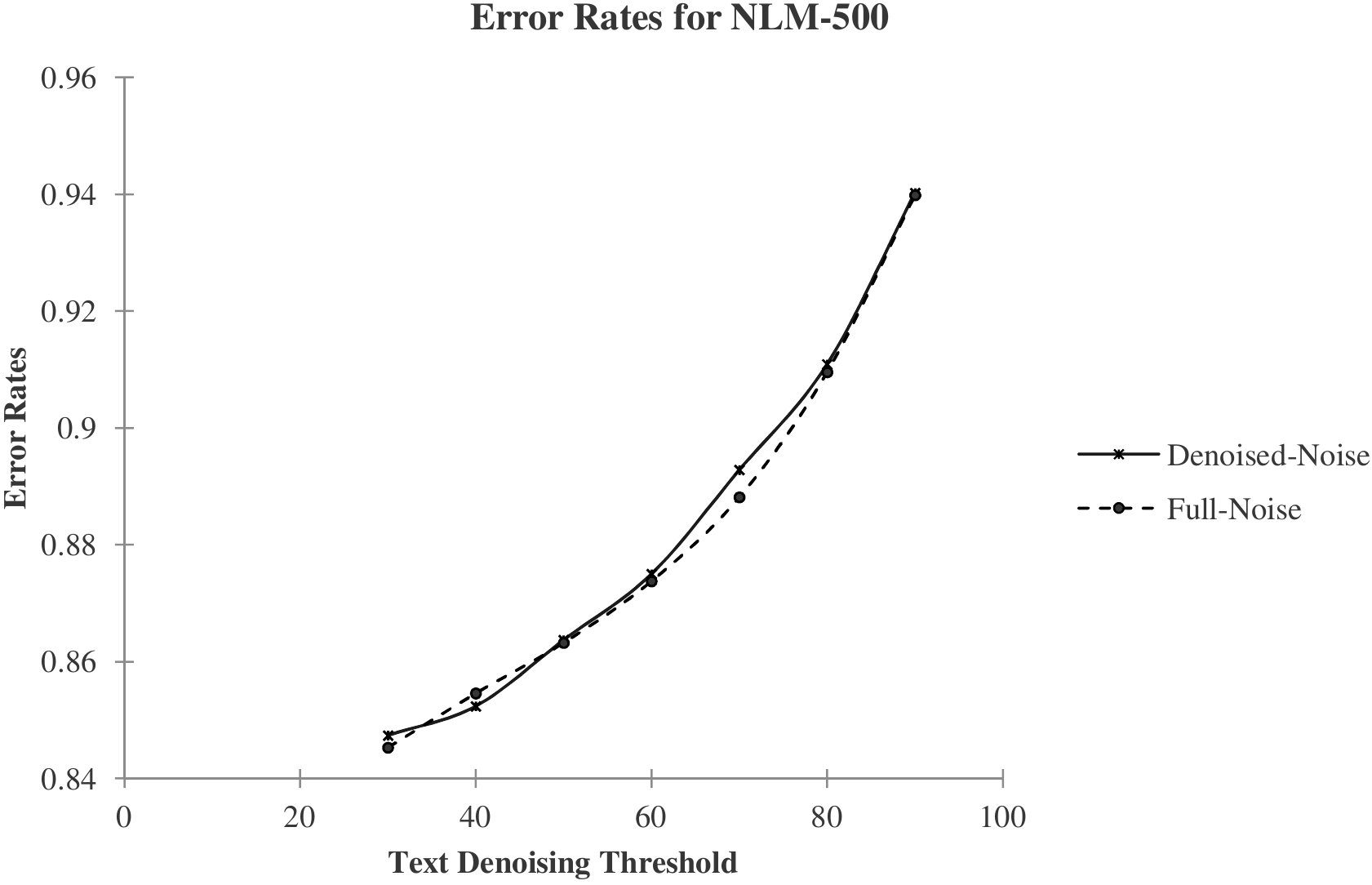}}
  \caption{Text denoising threshold for NLM-500 dataset}
  \label{fig:nlmthreshold}
\end{figure*}

%%%%%%%%%%%%%%%%%%%%%%%%%%%%%%%%%%%%%%%%%%%%%%%%%%%%%%%%%%%%%%%%%%%%%%%%%%%%%%%%%%%%%%%%%%%%%%%%%%%%%%%%%%%%%%%%%%%%%%%%%%%%%%%%%%%%%%%%%%%%%%%%%%%%%%%%%%%%%%%%
\subsection{Text Denoising Threshold}
\label{threshold}
%%%%%%%%%%%%%%%%%%%%%%%%%%%%%%%%%%%%%%%%%%%%%%%%%%%%%%%%%%%%%%%%%%%%%%%%%%%%%%%%%%%%%%%%%%%%%%%%%%%%%%%%%%%%%%%%%%%%%%%%%%%%%%%%%%%%%%%%%%%%%%%%%%%%%%%%%%%%%%%%
To find the appropriate text denoising threshold for keyphrase indexing, we evaluate Maui's performance on each dataset by increasing the text denoising threshold in increments of 10\% from 30\% to 90\%. As we vary the threshold, we plot the error rates of Maui on different test sets. Because Maui applies a supervised learning algorithm to develop its trained models, the best-fitted model should be where the test error has its global minimum. Therefore, the objective of this plotting is to discover the global minimum with its Full-text and Denoised-text trained models. This global minimum will  eventually be the denoising threshold. 

As an example, Figure \ref{fig:faothreshold} shows the error rates for different denoising thresholds. It is notable that when we use the Denoised-text trained models to extract keyphrases from either of the test sets, or use the Full-text trained models on Denoised-text test sets, Maui has its global minimum at 70\% denoising (Figure \ref{fig:faothreshold1}). From this point on, the error rate increases and thus indicates an overfitting in Maui's models. Figure \ref{fig:faothreshold2}, on the other hand, shows that no matter which trained model is used, full-text or denoised, the error rate for noise test sets increases with increasing thresholds. This indicates that noise texts are not content-rich as Maui fails to extract a substantial number of keyphrases from them. Figure \ref{fig:faothreshold} shows that Maui's best performing pair is \textit{Denoised-Full}---those models that are trained with denoised texts for keyphrase extraction from full texts. 

Similarly, Maui's best-fitted models with denoised texts for the CERN-290 dataset are also at the 70\% threshold (Figure \ref{fig:cernthreshold1}). Maui's models---full-text or denoised---experience overfitting after this threshold. Similar to what we observed for the FAO-780 dataset, Maui has no improvement with either of its trained models on the noise test set (Figure \ref{fig:cernthreshold2}). Maui best performs on the CERN-290 dataset with its full-text trained models to extract keyphrases from denoised texts (Figure \ref{fig:cernthreshold}), unlike that on FAO-780.

Maui's best-fitted models with denoised texts for the NLM-500 corpus are also at 70\% threshold, except for the full-text trained model on denoised-text test sets,  (Figure \ref{fig:nlmthreshold1}). In fact, Maui does not have any global minimum when it applies a full-text trained model on denoised test sets until its denoising threshold is set at 90\%. Maui's performance with noise texts from this domain is similar to that from the other domains (Figure \ref{fig:nlmthreshold2}). Like its performance on FAO-780 dataset, Maui best performs on full text test sets with its denoised text trained model for NLM-500.

These observations lead us to set the denoising threshold at 70\%. At this threshold, Maui predicts keyphrases from unseen test examples most accurately. %%%% I DON'T UNDERSTAND THE NEXT PART OF THE SENTENCE and after reaching this point, Maui's models become complex which can predict keyphrases from training examples but from unseen test examples.

\begin{table*}[ht]
\centering
\subfloat[Maui's performance on FAO-780 dataset with 70\% of the texts]{
\begin{tabular}{|c|c|c|c|c|c|c|c|c|}
\hline 
\multicolumn{5}{|c|}{} & \multicolumn{3}{c|}{\textbf{Benchmark Performance}} & \tabularnewline
\hline 
\multirow{1}{*}{\textbf{Trained Model}} & \textbf{Test Set} & \textbf{Precision} & \textbf{Recall} & \textbf{F-score} & \textbf{Precision} & \textbf{Recall} & \textbf{F-score} & \textbf{\textit{t} value}\tabularnewline
\hline 
Denoised Text & Denoised Text & 30.02 & 32.92 & 31.36 & \multirow{3}{*}{30.56 } & \multirow{3}{*}{33.47} & \multirow{3}{*}{31.86} & 2.23\tabularnewline
\cline{1-5} \cline{9-9} 
\textbf{Denoised Text} & \textbf{Full Text} & \textbf{30.49}  & \textbf{33.50} & \textbf{31.87}  &  &  &  & \textbf{2.76}\tabularnewline
\cline{1-5} \cline{9-9} 
Full Text & Denoised Text & 30.48 & 32.96  & 31.63  &  &  &  & 1.81\tabularnewline
\hline 
\end{tabular}
\label{faoprf}}

\qquad

\subfloat[Maui's performance on CERN-290 dataset with 70\% of the texts]{
\begin{tabular}{|c|c|c|c|c|c|c|c|c|}
\hline 
\multicolumn{5}{|c|}{} & \multicolumn{3}{c|}{\textbf{Benchmark Performance}} & \tabularnewline
\hline 
\multirow{1}{*}{\textbf{Trained Model}} & \textbf{Test Set} & \textbf{Precision} & \textbf{Recall} & \textbf{F-score} & \textbf{Precision} & \textbf{Recall} & \textbf{F-score} & \textbf{\textit{t} value}\tabularnewline
\hline 
Denoised Text & Denoised Text & 24.38  & 25.33  & 24.79 & \multirow{3}{*}{24.58} & \multirow{3}{*}{25.56} & \multirow{3}{*}{24.99} & 2.16\tabularnewline
\cline{1-5} \cline{9-9} 
Denoised Text & Full Text & 23.99  & 24.95 & 24.42 &  &  &  & 2.26\tabularnewline
\cline{1-5} \cline{9-9} 
Full Text & Denoised Text & 24.38 & 25.40 & 24.82 &  &  &  & 1.31\tabularnewline
\hline 
\end{tabular}
\label{cernprf}}

\qquad

\subfloat[Maui's performance on NLM-500 dataset with 70\% of the texts]{
\begin{tabular}{|c|c|c|c|c|c|c|c|c|}
\hline 
\multicolumn{5}{|c|}{} & \multicolumn{3}{c|}{\textbf{Benchmark Performance}} & \tabularnewline
\hline 
\multirow{1}{*}{\textbf{Trained Model}} & \textbf{Test Set} & \textbf{Precision} & \textbf{Recall} & \textbf{F-score} & \textbf{Precision} & \textbf{Recall} & \textbf{F-score} & \textbf{\textit{t} value}\tabularnewline
\hline 
Denoised Text & Denoised Text & 29.14 & 32.36 & 30.66 & \multirow{3}{*}{29.69 } & \multirow{3}{*}{32.74} & \multirow{3}{*}{31.13} & 2.01\tabularnewline
\cline{1-5} \cline{9-9} 
\textbf{Denoised Text} & \textbf{Full Text} & \textbf{29.96} & \textbf{33.22} & \textbf{31.50} &  &  &  & \textbf{3.52}\tabularnewline
\cline{1-5} \cline{9-9} 
Full Text & Denoised Text & 28.99 & 32.00 & 30.40 &  &  &  & 1.85\tabularnewline
\hline 
\end{tabular}
\label{nlmprf}}

\caption{Precision, recall and F-score of Maui with text denoising}
\label{mauiprf}
\end{table*}
\begin{table*}[ht]
\centering

\subfloat[Maui's indexing agreements on FAO-780 dataset with 70\% of the texts]{
\begin{tabular}{|c|c|c|c|c|c|c|c|}
\hline 
\multicolumn{5}{|c|}{} & \multicolumn{3}{c|}{\textbf{Benchmark Performance}}\tabularnewline
\hline 
\multirow{1}{*}{\textbf{Trained Model}} & \textbf{Test Set} & \textbf{Hooper} & \textbf{Rolling} & \textbf{Cosine} & \textbf{Hooper} & \textbf{Rolling} & \textbf{Cosine}\tabularnewline
\hline 
Denoised Text & Denoised Text & 0.18 & 0.29 & 0.30 & \multirow{3}{*}{0.18} & \multirow{3}{*}{0.30} & \multirow{3}{*}{0.31}\tabularnewline
\cline{1-5} 
\textbf{Denoised Text} & \textbf{Full Text} & \textbf{0.18}  & \textbf{0.30} & \textbf{0.31} &  &  & \tabularnewline
\cline{1-5} 
Full Text & Denoised Text & 0.18 & 0.30 & 0.30 &  &  & \tabularnewline
\hline 
\end{tabular}
\label{faoiic}}

\qquad

\subfloat[Maui's indexing agreements on CERN-290 dataset with 70\% of the texts]{
\begin{tabular}{|c|c|c|c|c|c|c|c|}
\hline 
\multicolumn{5}{|c|}{} & \multicolumn{3}{c|}{\textbf{Benchmark Performance}}\tabularnewline
\hline 
\multirow{1}{*}{\textbf{Trained Model}} & \textbf{Test Set} & \textbf{Hooper} & \textbf{Rolling} & \textbf{Cosine} & \textbf{Hooper} & \textbf{Rolling} & \textbf{Cosine}\tabularnewline
\hline 
Denoised Text & Denoised Text & 0.14 & 0.24  & 0.24  & \multirow{3}{*}{0.14 } & \multirow{3}{*}{0.24} & \multirow{3}{*}{0.24}\tabularnewline
\cline{1-5} 
Denoised Text & Full Text & 0.14 & 0.24  & 0.24 &  &  & \tabularnewline
\cline{1-5} 
Full Text & Denoised Text & 0.14 & 0.24 & 0.24 &  &  & \tabularnewline
\hline 
\end{tabular}
\label{cerniic}}

\qquad

\subfloat[Maui's indexing agreements on NLM-500 dataset with 70\% of the texts]{
\begin{tabular}{|c|c|c|c|c|c|c|c|}
\hline 
\multicolumn{5}{|c|}{} & \multicolumn{3}{c|}{\textbf{Benchmark Performance}}\tabularnewline
\hline 
\multirow{1}{*}{\textbf{Trained Model}} & \textbf{Test Set} & \textbf{Hooper} & \textbf{Rolling} & \textbf{Cosine} & \textbf{Hooper} & \textbf{Rolling} & \textbf{Cosine}\tabularnewline
\hline 
Denoised Text & Denoised Text & 0.18 & 0.30 & 0.30 & \multirow{3}{*}{0.18} & \multirow{3}{*}{0.30} & \multirow{3}{*}{0.31}\tabularnewline
\cline{1-5} 
\textbf{Denoised Text} & \textbf{Full Text} & \textbf{0.19} & \textbf{0.31} & \textbf{0.31} &  &  & \tabularnewline
\cline{1-5} 
Full Text & Denoised Text & 0.18 & 0.30 & 0.30 &  &  & \tabularnewline
\hline 
\end{tabular}
\label{nlmiic}}

\caption{Inter-indexing agreements of Maui with text denoising}
\label{mauiiic}
\end{table*}

%%%%%%%%%%%%%%%%%%%%%%%%%%%%%%%%%%%%%%%%%%%%%%%%%%%%%%%%%%%%%%%%%%%%%%%%%%%%%%%%%%%%%%%%%%%%%%%%%%%%%%%%%%%%%%%%%%%%%%%%%%%%%%%%%%%%%%%%%%%%%%%%%%%%%%%%%%%%%%%%
\section{Results and Discussions}
\label{results}
%%%%%%%%%%%%%%%%%%%%%%%%%%%%%%%%%%%%%%%%%%%%%%%%%%%%%%%%%%%%%%%%%%%%%%%%%%%%%%%%%%%%%%%%%%%%%%%%%%%%%%%%%%%%%%%%%%%%%%%%%%%%%%%%%%%%%%%%%%%%%%%%%%%%%%%%%%%%%%%%
In this section, we discuss the performance of Maui with text denoising and compare this with its benchmark performance. 

Table \ref{faoprf} shows the precision, recall and F-score of Maui with denoised texts and its benchmark performance on the FAO-780 dataset. Maui, as it uses its denoised text and full-text trained models on denoised text test sets, achieves F-scores of $31.36$ and $31.63$, respectively, compared to its benchmark F-score of $31.86$. By applying the 10-fold cross validation \textit{t}-test, we see that for these two cases, the \textit{t}-values are $2.23$ and $1.81$, respectively, which means that the differences between the F-scores are not statistically significant at $\alpha$ = $0.05$. In other words, the benchmark performance of Maui cannot be said to be different than that with text denoising. On the other hand, Maui's F-score with its denoised text trained model on full-text keyphrase extraction is $31.87$. A significance test shows that its \textit{t}-value is $2.76$ which indicates that it is different at a significance level of $\alpha$ = $0.02$. So, with 98\% confidence we can say that the result is better than the benchmark performance. In addition, from Table \ref{faoiic}, we can see that Maui's agreements with the gold standards are as good as the benchmark agreements. This demonstrates that the indexing quality of Maui has not been compromised with text denoising.

For CERN-290 dataset, although Maui could not outperform its benchmark F-score of $24.99$, none of the \textit{t}-values are significant at $\alpha$ = $0.05$. In other words, its performance with denoised texts cannot be said to be different than its benchmark performance with a 95\% confidence level. Its best performance with denoised texts is when it uses the full-text trained model to extract keyphrases from denoised texts (F-score of $24.82$). Maui's performance details on the CERN-290 dataset are given in Table \ref{cernprf}. 

It is noteworthy that the performance of BibClassify, a specialized keyphrase indexer developed by CERN for physics documents, on the CERN-290 documents is $15.40$ precision, $24.3$ recall and $18.80$ F-score~\cite{Olena:2009}. If we compare this performance with Maui (Table \ref{cernprf}), then we can see that using 70\% of the text, Maui outperforms BibClassify.

Interestingly enough, although Maui agrees less with the gold standard keyphrases for CERN-290 dataset than that for FAO-780, its agreements on keyphrases with denoised texts are as good as its benchmark performance. The details for its inter-indexing agreement measures on the CERN-290 dataset are listed in Table \ref{cerniic}. Like its performance on FAO-780, we see that Maui extracts quality keyphrases from physics documents when paired with text denoising compared to that with full texts. 

Maui's best performance for the NLM-500 corpus is with a denoised text trained model on full texts. Its F-score of $31.50$ outperforms the benchmark F-score of $31.13$ at the significance level of $\alpha$ = $0.05$. However, although its other two F-scores with denoised texts is somewhat lower than its benchmark F-score, they are not statistically significant with \textit{t}-values of $2.01$ and $1.85$ (Table \ref{nlmprf}). Maui's inter-indexing agreement on NLM-500 is somewhat similar to that on FAO-780 except that it agrees more with NLM-500 gold standards than its benchmark performance (Table \ref{nlmiic}).

%%%%%%%%%%%%%%%%%%%%%%%%%%%%%%%%%%%%%%%%%%%%%%%%%%%%%%%%%%%%%%%%%%%%%%%%%%%%%%%%%%%%%%%%%%%%%%%%%%%%%%%%%%%%%%%%%%%%%%%%%%%%%%%%%%%%%%%%%%%%%%%%%%%%%%%%%%%%%%%%
\section{Conclusions}
\label{conclusions}
%%%%%%%%%%%%%%%%%%%%%%%%%%%%%%%%%%%%%%%%%%%%%%%%%%%%%%%%%%%%%%%%%%%%%%%%%%%%%%%%%%%%%%%%%%%%%%%%%%%%%%%%%%%%%%%%%%%%%%%%%%%%%%%%%%%%%%%%%%%%%%%%%%%%%%%%%%%%%%%%
In this experiment, we consider full texts as well as their denoised and noise parts from different domains like food and agriculture, physics, and biomedical science. For each genre of texts, we have seen that Maui's trained models overfit if we set denoising threshold beyond 70\%. Considering this as our denoising threshold for keyphrase indexing, we test Maui on these texts, full and reduced.

From its experimental results, we show in this paper that text denoising improves Maui's performance for the biomedical science texts, or it allows Maui to perform as good as its benchmark performance on the food and agriculture, and the physics texts. It does so by reducing Maui's training and test sets to 70\%.  For instance, the FAO-780 dataset of $24$ million words has been reduced by text denoising to a set of $17$ million words and Maui does not perform poorer than its benchmark performance. In other words, the $7$ million removed words are not potential candidates as keyphrases. Text denoising, as expected, left out the words from being considered so. 

Although there are some cases where Maui, when paired with text denoising, experiences marginally lower F-score than its benchmark, indexing agreement measures show that its indexing quality has never been compromised; it extracts even better quality keyphrases from biomedical texts than its benchmark.  

It is noteworthy that during this experiment, we did not change the way Maui works rather we tested it to see its response on a set of reduced texts.

Therefore, our experimental findings reveal that
\begin{itemize}[leftmargin=*]
\item document size, per se, does not have the suggested effect on keyphrase indexing---it is the content richness that plays the key role in indexing,
\item text denoising produces a content-rich set of sentences which can improve indexer performance,
\item the noise texts, i.e.,\ the removed text, do not improve indexing rather they increase the error rates,
\item text denoising is useful not only for biomedical relation extraction but also for keyphrase indexing, and
\item text denoising can be used for different domains other than biomedical science
\end{itemize}

With these results in mind and recalling that there are other indexers that use different features to train their machine learning models, we are interested in further
investigating the effect of text denoising on them. In addition, when paired with text denoising, Maui performs better on biomedical texts than texts from agriculture and physics. Because it has been originally developed for relation mining in biomedical texts, we are also interested to explore the reasons behind the success of text denoising for keyphrase indexing in this domain. These investigations are left for future work.

%\newpage
%%%%%%%%%%%%%%%%%%%%%%%%%%%%%%%%%%%%%%%%%%%%%%%%%%%%%%%%%%%%%%%%%%%%%%%%%%%%%%%%%%%%%%%%%%%%%%%%%%%%%%%%%%%%%%%%%%%%%%%%%%%%%%%%%%%%%%%%%%%%%%%%%%%%%%%%%%%%%%%%
\section{Acknowledgments}
%%%%%%%%%%%%%%%%%%%%%%%%%%%%%%%%%%%%%%%%%%%%%%%%%%%%%%%%%%%%%%%%%%%%%%%%%%%%%%%%%%%%%%%%%%%%%%%%%%%%%%%%%%%%%%%%%%%%%%%%%%%%%%%%%%%%%%%%%%%%%%%%%%%%%%%%%%%%%%%%
This work was partially funded through a Natural Sciences and Engineering Research Council of Canada (NSERC) Discovery Grant to Robert E. Mercer. Thanks are due to Olena Medelyan for providing the datasets and Maui-- the indexer and her time to time responses regarding our queries. We also acknowledge Ian H. Witten for his experienced remarks on experiment setup and results.

\bibliographystyle{abbrv}
\bibliography{rushdibibliography}

\end{document}